\documentclass[runningheads]{llncs}

\usepackage{graphicx}
\usepackage{csquotes}
\usepackage[backend=biber, style=numeric, sorting=none]{biblatex}
\begin{document}

\title{Design Your Life: User-Initiated Design of Technology to Support Independent Living of Young Autistic Adults}

\titlerunning{Design Your Life: User-Initiated Design for Young Autistic Adults}

\author{Thijs Waardenburg\inst{1,2}\orcidID{0000-0002-4358-833X} \and
Niels van Huizen\inst{2} \and
Jelle van Dijk\inst{2}\orcidID{0000-0001-8224-3302} \and
Maurice Magnée\inst{1}\orcidID{0000-0003-2423-2132} \and
Wouter Staal\inst{3,4,5}\orcidID{0000-0002-3276-1133} \and
Jan-Pieter Teunisse\inst{1,6} \and
Mascha van der Voort\inst{2}\orcidID{0000-0002-4800-7557}
}

\authorrunning{T. Waardenburg et al.}

\institute{HAN University of Applied Sciences, Nijmegen, The Netherlands \and
University of Twente, Enschede, The Netherlands \and 
Karakter Child and Adolescent Psychiatry University Centre, Nijmegen, The Netherlands. \and
Radboud University Nijmegen Medical Centre Nijmegen, The Netherlands \and
Leiden University, Leiden, The Netherlands \and
Dr. Leo Kannerhuis, Oosterbeek, The Netherlands \\
\email{thijs.waardenburg@han.nl, j.c.vanhuizen@utwente.nl, jelle.vandijk@utwente.nl, maurice.magnee@han.nl, w.staal@karakter.com, janpieter.teunisse@han.nl, m.c.vandervoort@utwente.nl}
}

\maketitle           

\begin{abstract}
This paper describes the development of and first experiences with ‘Design Your Life’: a novel method aimed at user-initiated design of technologies supporting young autistic adults in independent living. A conceptual, phenomenological background resulting in four core principles is described. Taking a practice-oriented Research-through-Design approach, three co-design case studies were conducted, in which promising methods from the co-design literature with the lived experiences and practical contexts of autistic young adults and their caregivers is contrasted. This explorative inquiry provided some first insights into several design directions of the Design Your Life-process. In a series of new case studies that shall follow, the Design Your Life-method will be iteratively developed, refined and ultimately validated in practice.

\keywords{Autism  \and Assistive Technology \and Phenomenology \and Participatory Design \and Research-through-Design \and Independent Living.}
\end{abstract}

\section{Introduction}
An estimated one in hundred people worldwide have been diagnosed with autism \cite{idring_autism_2012}. While many autistic people have both the motivation and intellectual capacity to contribute fully to society, the group shows a relatively large percentage of school dropout and unemployment, especially if a suitable support network is lacking \cite{scheeren_research_2015}. In recent years, participatory design has become of interest in the context of designing technologies that could support autistic people in daily life \cite{newbutt_using_2016, wilson_co-design_2019}. The motives for adopting a participatory design approach range from “addressing a pragmatic need to increase the fit between features and users’ requirements” to “idealistic agendas related to empower people, democratise innovation and designing alternative futures” \cite{frauenberger_blending_2017}. \par \noindent
Morally, a call for empowerment through participation has been advocated more broadly within what is called the ‘neurodiversity movement’ \cite{fletcher-watson_making_2019}. According to Frauenberger, Makhaeva and Spiel, there is “consensus that participatory design is particularly powerful when creating technologies for groups who are typically marginalised in design and have life-worlds which are far removed from those of designers and researchers” \cite{frauenberger_blending_2017}. \par 
The epistemic value of participatory design has been invoked as well: without a proper understanding of autistic experiences, designers may be at risk to end up creating technologies that are both stigmatising and ineffective \cite{fletcher-watson_autism_2019}. Yet, evidence for the effectiveness of technology-based interventions remains limited \cite{zervogianni_framework_2020, frauenberger_designing_2016}. One reason may be that it is inherently difficult for (non-autistic) designers to empathise with the lived experience of autism. Incorporating one’s lived, subjective experiences into the design process is difficult, but this is especially complex in the context of autism due to differences in perceptions between autistic and non-autistic people. This is something Milton refers to as the ‘double empathy problem’ \cite{milton_ontological_2012}. Another reason may be that the autistic population is itself highly heterogeneous, with each person having both individual support needs as well as highly personal and sometimes quite specific interests and capabilities. Estimates of intellectual capacities are highly variable in the autism spectrum as well, varying from intellectual disability in about forty percent of the autistic population to normal and (very) high IQ ranges \cite{idring_autism_2012}. \par
Finally, many assistive technologies tend to be developed within the context of professionals and health care organisations seeking possibilities for technologically augmenting (or even partly replacing) existing therapeutic methods. This means that the function of the resulting living space, device or app will be grounded in the logic of formal therapy structures and objectives. As a result, such technologies may come to reflect a frame in which the autistic person is at the core seen as a person who displays dysfunctional behaviour, which needs somehow to be corrected, with the ‘healthy’ neurotypical person as the target model. In such cases, it can be argued that the main user of the product is in some sense not the autistic person but, for instance, the therapist, who uses the product to bring about a desired normative change in the patient. It is precisely this medical frame that autistic people may find disempowering. \par
Starting with these challenges in mind, a first iteration of the ‘Design Your Life-method’ (hereafter DYL) is developed: a co-design method that helps young autistic adults (hereafter YAA\footnote{There appears to be no clear consensus on the naming of 'a person with an autism spectrum disorder' \cite{kenny_which_2016}. There are several variations, for instance: ‘autistic person', 'person with autism', 'person who has autism', ‘person on the spectrum’, etc. In this research, the term ‘young autistic adult’ (abbr. as ‘YAA’, also in plural form) is used, without the intention to disregard various conceptions of the designation. Usually, 'young adult' refers to an age category of 16-30 years. In this study, a broader definition is used, namely 16-35 years, because the development towards more independence may occur on a later age.}) and their caregivers in designing a personalised, supportive, technological environment that contributes to independent living. Its purpose is two-fold. First, DYL brings creative forms of shared sense-making to the care practice. Second, DYL signifies the next step in reaching autistic empowerment by providing autistic individuals not just participation in a designer’s project, but to instead provide them with tools to design and implement their own supportive interventions. This is guided by a phenomenological perspective on autistic experience as being fundamentally embodied, holistic and contextual in nature \cite{frith_enactive_2003, de_jaegher_embodiment_2013}. \par 
The rest of this article is organised as follows: With the challenges and values described in this introduction in mind, four core principles are described. These were brought in practice and further refined during three case studies described in the subsequent chapter. Reflecting on the cases studies, a first version of the DYL process is developed. Conclusions are discussed in the remainder of the article.

\section{Four Core Principles}
DYL underscores the phenomenological nature of (autistic) experiences. At its essence, phenomenology is the philosophical study of experience. It premiered in the transcendental phenomenology of German philosopher Edmund Husserl. Dourish explains that "Husserl was frustrated by the idea that science and mathematics were increasingly conducted on an abstract plane that was disconnected from human experience and human understanding” \cite{dourish_where_2001}. In other words, Husserl believed that abstract reasoning and idealised conceptualisation of world phenomena would severely gloss over everyday experiences. \par
Phenomenology is not new to the field of human-centred design. In the 1980s, a study by Winograd and Flores \cite{winograd_understanding_1986} invoked the work of phenomenologist Martin Heidegger to shed a new light on human-computer interaction. Later, Weiser \cite{weiser_computer_1991} would build upon phenomenological insights to explore his vision for ubiquitous computing. Dourish places endeavours as such in a historical perspective. Ac-cording to him, the very emergence of tangible and social computing signifies an appreciation of phenomenological thought. He writes:
\begin{displayquote} Instead of drawing on artefacts in the everyday world, it draws on the way the everyday world works or, perhaps more accurately, the way we experience the world. Both approaches draw on the fact that the ways in which we experience the world are through directly interacting with it, and that we act in the world by exploring the opportunities for action that it provides to us – whether through its physical configuration, or through socially constructed meanings. In other words, they share an understanding that you cannot separate the individual from the world in which that individual lives and acts \cite{dourish_where_2001}. \end{displayquote} \par \noindent
Inspired by this phenomenological background, the four core principles for DYL are presented.

\subsection{Focus on Experience}
Although it is already entailed by a phenomenological approach itself, experience is the focal point of DYL. In the context of autism, one must observe that the design challenge is too often described in terms of functional and psycho-social limitations. Frauenberger, Makhaeva and Spiel explain that such a "reductionist model of disabilities” glosses over the importance of experience that autistic individuals have with technologies \cite{husserl_crisis_2006}. A mismatch ensues in which autistic individuals may abandon technologies that they find ineffective. This observation comes close to Husserl's aforementioned criticism of abstraction. By describing autism in terms of ‘technical limitations’, one may expect that many technologies related to autism come in the shape of a ‘technological fix'. Or, to put it in words of Husserl: “Merely fact-minded sciences make merely fact-minded people” \cite{husserl_crisis_2006}. In light of technology abandonment, better incorporation of autistic experiences is one of the core principles of DYL. To this end, DYL builds upon phenomenology as the philosophical study of experience to comprehend and articulate autistic experiences as valuable input in the design process. In this regard, phenomenologists have emphasised the cognitive, social, cultural and biological constituents of a user experience \cite{frith_enactive_2003, sarmiento-pelayo_co-design_2015, svanaes_interaction_2013}. These constituents must be well-understood before a technology can become successfully integrated in autistic life-worlds. To stay in phenomenological terminology: the technology must become properly ‘embodied’.

\subsection{Action-Oriented Tinkering}
DYL encourages ‘action-oriented tinkering’. This builds on what Svanæs \cite{svanaes_interaction_2013} calls ‘the feel dimension’: “[in] the same way as you see a rose, not a collection of petals, and hear a musical theme, not a sequence of notes, you perceive the interactive behaviour of an interactive artefact not as a collection of action/reaction pairs, but as a meaningful interactive whole”. In this regard, Svanæs argues that it would be nonsensical trying to articulate an experience in terms of logical analysis. Rather, an experience must be lived through; it must be tried and felt. Its implications for design are clear: prototypes must evolve ‘on stage’ by acting out future scenarios rather than “defining design as a separate activity” \cite{svanaes_interaction_2013}. Only then, Svanæs claims, the lived experience of the user is properly reflected in the design outcome.

\subsection{User-Initiated Design}
DYL promotes user-initiated design (UID). This is in contrast with most co-design methods that do involve stakeholders during the design of a product or service, but do not equip them with the tools to design own, specific solutions \cite{sarmiento-pelayo_co-design_2015}. Choosing your own technology means that the design process departs from your own lived experience with technology. This may represent a state of empowerment in its own right. UID steps away from the concept of ‘one-size-fits-all’ solutions that are designed for the masses.

\subsection{Off-the-shelf Technologies}
To enable user-initiated design, DYL promotes using already available technologies, ranging from non-digital low-tech to digital high-tech, instead of developing new (‘assistive’) technologies. The reason for this is two-fold. On one hand, generating design ideas proved to be challenging as YAA have difficulties with conceptualising a concrete, technological intervention ‘out of nothing’. By introducing off-the-shelf technologies already at the start of the design process, it is easier to envisage the design possibilities. A more practical consideration is that off-the-shelf technologies are relatively low-cost and “present opportunities for applications that can be individualised at lower costs than using the more traditional custom hardware solutions” \cite{hayes_using_2013}.

\section{Case Studies}
Three case studies were conducted between March and September 2020, following the tradition of ‘Research-through-Design’, focusing not so much on the end result but on insights gained during the design practice \cite{zimmerman_research_2007}. These case studies were conducted for this study to explore several design directions of the underlying DYL process: the steps to be taken during the design of solutions. \par
All case studies involved working with stakeholders embedded in a real care practice and consisted of designing and trialling a prototype of a concrete DYL toolkit. This toolkit represented the underlying DYL process and provided hands-on guidance for a YAA and caregiver to move through the design process one or more times (so-called ‘iterations’). The prototypes were designed, developed and tested by Industrial Design Engineering students from the University of Twente. In all cases, Stanford’s d.school design thinking process was used as a basis. This model assumes five phases, which can be repeated in various orders: Empathize (i.e. observing, engaging, immersing), Define (i.e. understanding a useful challenge), Ideate (i.e. exploring solutions), Prototype (i.e. realising ideas) and Test (i.e. gathering feedback, refining solutions) (Figure 1) \cite{doorley_design_2018}.
\begin{figure}
\includegraphics[width=\textwidth]{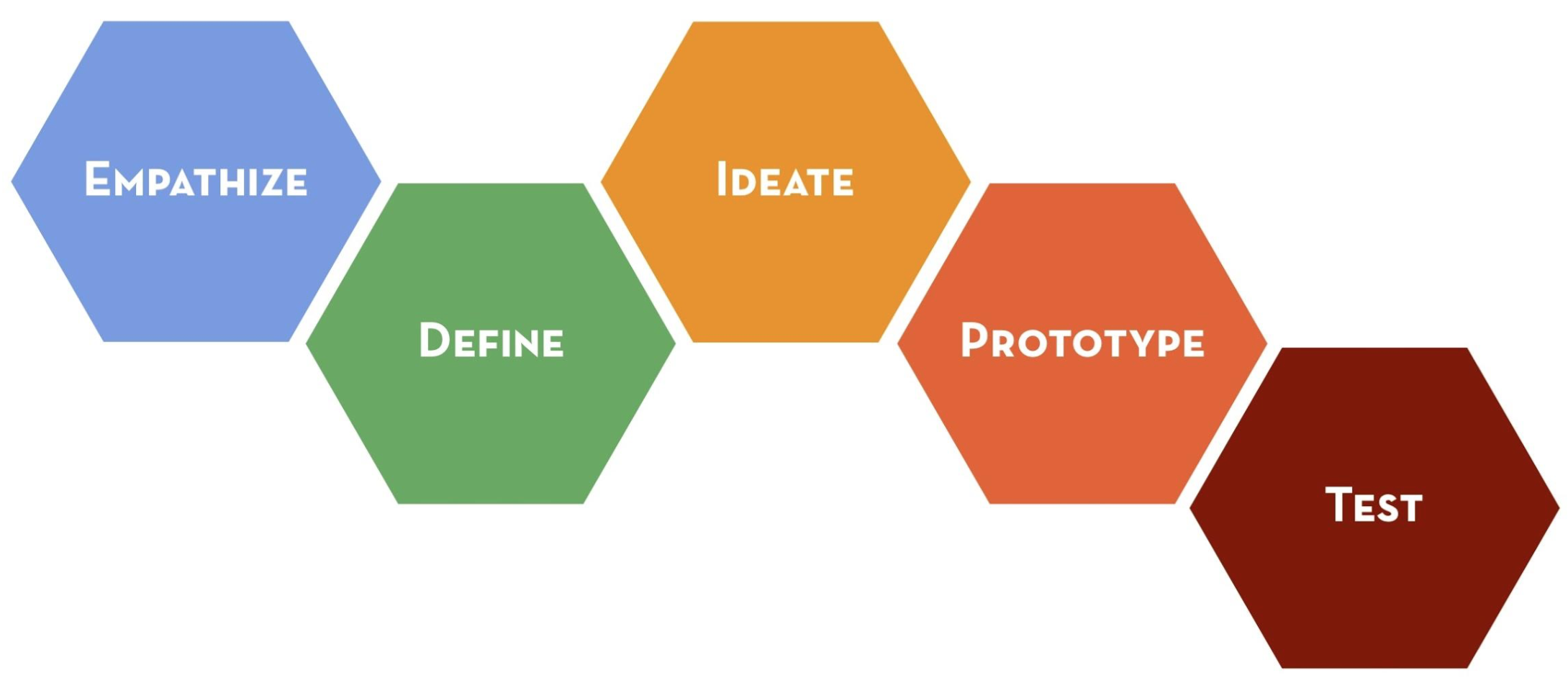}
\caption{Stanford’s d.school design thinking process \cite{doorley_design_2018}, CC BY-NC-SA 4.0.} \label{fig1}
\end{figure}

For each phase it was further explored how to support the YAA and caregiver with concrete tools (instruction cards, visual materials, expressive tools, a representation of state and process and so on) to move through this basic process, and how to concretely give form to this phase. This meant on the one hand selecting and adapting various promising tools and techniques from the co-design literature as well as getting to know the YAA and caregiver that were participating in the case, being informed and inspired by their lived experiences and practices, and subsequently bringing theory and practice together in concrete design choices for the method and toolkit. \par

\subsection{First Case Study}
The first case study involved Tim\footnote{To protect the privacy of participants, pseudonyms are used.}, a 14-year-old adolescent and his mother (Tim lives at home). Although Tim is not a young adult, it was chosen to carry out a case study with him, mainly because the development to independence begins at a younger age than adulthood. Tim’s mother feels responsible for the development in independence of her son. However, Tim seems to find that less urgent. For instance, at one moment he remarked something along the lines of ‘Why should I learn to tie my laces? My mother is much better at it than I am’. Also, “[i]f he wants to do something, he wants someone to be with him” \cite{van_den_berg_designing_2020}. \par
Three different prototype forms were explored: tangible, digital and a tangible-digital combination. Although tangible co-design tools have been argued to provide a shared space for sense making and open up a rich freedom of expression for co-design participants, in this case Tim actually preferred the digital form, due to his affinity with ICT and his reluctance to write with a pen. Eventually an interactive PowerPoint prototype was developed and tested (Figure 2). It offers a step-by-step guidance in designing a solution. The various phases were translated to fit the life world of Tim and his mother. The ‘Empathize’ phase, for instance, was translated to ‘Discover who you are’ and ‘Prototype’ was translated to ‘Make your tool’. This prototype did not appeal sufficiently. For instance, it turned out to be too textual and too much effort to fill in all the text boxes. Commitment declined and no tool was designed by Tim and his mother (however, as said, this was not the primary goal of these case studies).
\begin{figure}
\includegraphics[width=\textwidth]{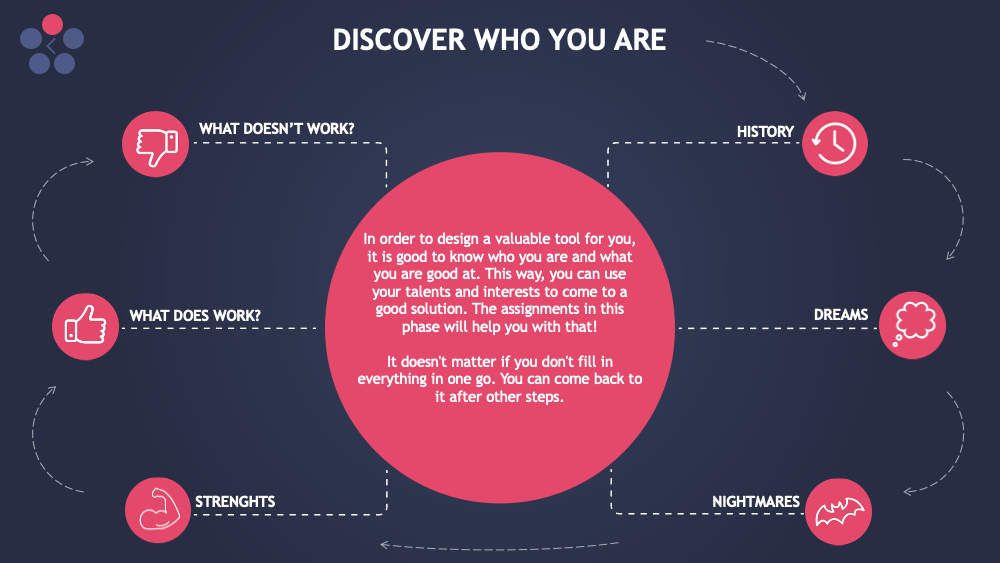}
\caption{One of the phases of an interactive prototype made with PowerPoint \cite{van_den_berg_designing_2020}} \label{fig2}
\end{figure}

\subsection{Second Case Study}
The second case study involved Paul, a 33-year-old YAA, and his caregiver. Paul lives at a mental healthcare institution that is involved in this research project. Slowly but steadily, he is working towards more independence. Again, three different forms were explored: A game board, an adaptable online guide and a physical toolkit. Low-fi prototypes were used as input (‘probes’) for a co-design session with Paul and his caregiver. During this session, they interacted with the prototypes to discover preferences and difficulties in the designs. It resulted in a new prototype design that consists of a set of ‘prompt cards’ that is aimed at stimulating, inspiring and guiding the design process. “The cards spark creativity or provides the users with new prompts to follow when they stagnate in the design process” \cite{schipper_design_2020} (Figure 3). Due to time limitations of the case study, no tool was developed.  However, useful insights were gained that could possibly be translated into a useful solution. For example, they thought it would be useful if Paul could have a ‘little voice on his shoulder’ to help him every now and then. Recording, sending (by a caregiver) and playback of messages can provide such functionality \cite{bouck_promoting_2021}.
\begin{figure}
\includegraphics[width=\textwidth]{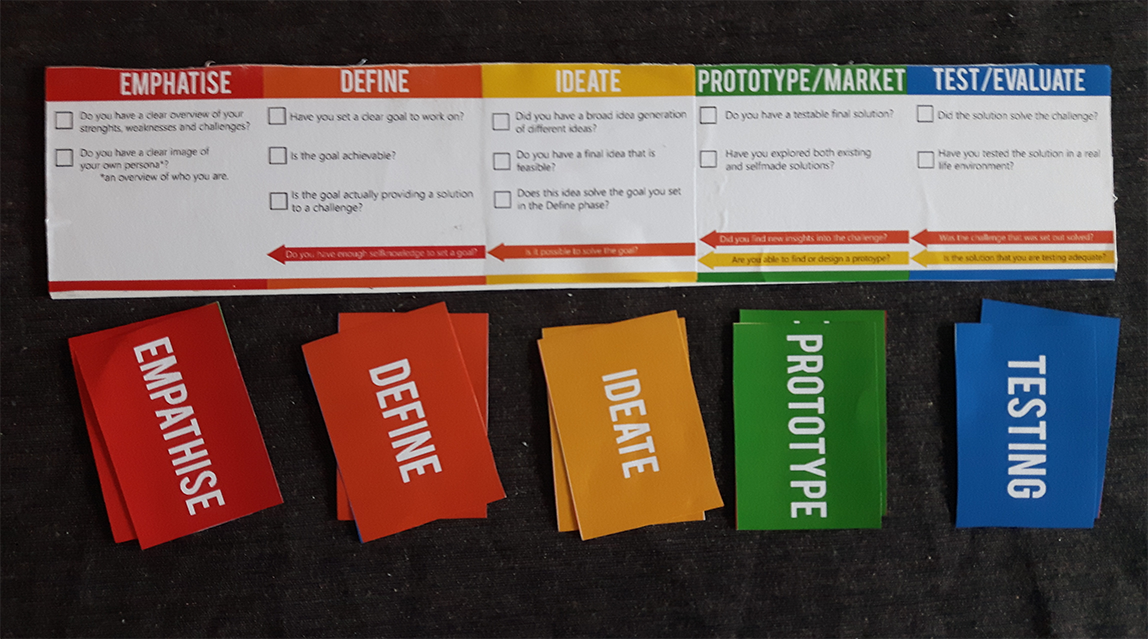}
\caption{A set of physical prompt cards to guide the design process \cite{schipper_design_2020}.} \label{fig3}
\end{figure}

\subsection{Third Case Study}
The third case study involved Vincent, a 23-year-old YAA, and his caregiver. Just as Paul, he lives at one of the involved mental healthcare institutions. In this case, a game-like prototype was developed (Figure 4). This really appealed to Vincent, because he likes to play board games. He had no trouble with the metaphors used. The prototype “consists out of six islands, five of which belong to the design thinking phases understand, define, ideate, prototype, and test. The final island includes the goal of the YAA” \cite{sagel_designing_2021}. The goal that they chose was to stimulate a better sleep-wake rhythm. Vincent’s caregiver already had a specific technology in mind beforehand: a care robot called Tessa (this robot reads out messages and agenda items). She hoped that this project would justify the purchase. How-ever, they found out that Vincent has a clear preference for non-auditory modalities (i.e. sounds). So, they ended up with a different, better fitting and simpler solution that was also much cheaper and easier to use: a wake-up light.
\begin{figure}
\includegraphics[width=\textwidth]{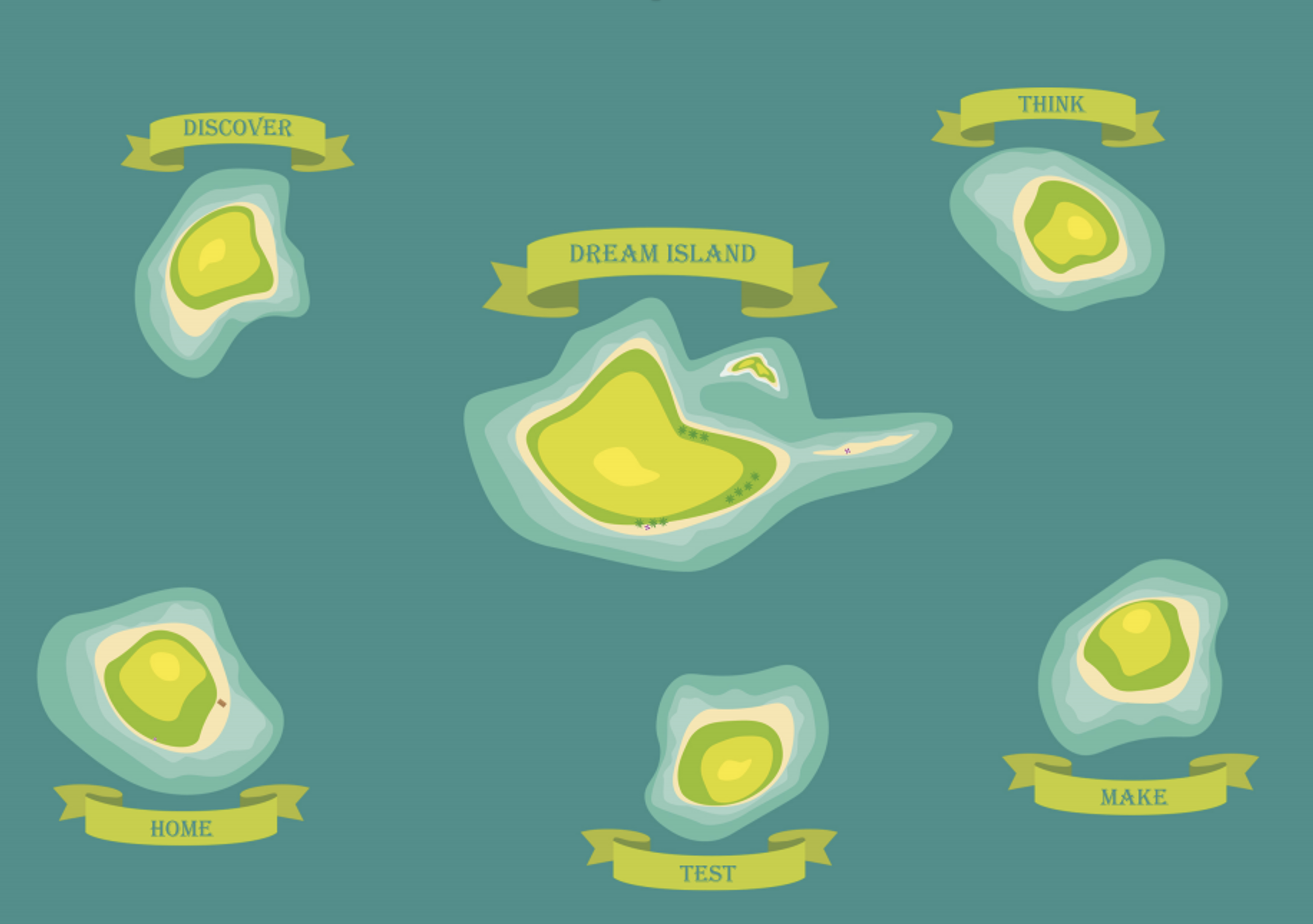}
\caption{A game-based prototype called “Good Trip!” \cite{sagel_designing_2021}.} \label{fig4}
\end{figure}

\section{Design Your Life Process}
The various proposed design processes of the three pilot studies were analysed and discussed with the involved researchers in an online Mural environment. This led to the first iteration of the DYL-process (Figure 5). Notably, these kinds of processes are not unique to participatory design projects: both in the digital and the tangible sphere, iterative processes with similar stages have become conventional to include stakeholders and make effective use of their expertise - also in the context of autism \cite{newbutt_using_2016, zervogianni_framework_2020}. The DYL-process and the four core principles together are called the DYL-method.
\begin{figure}
\includegraphics[width=\textwidth]{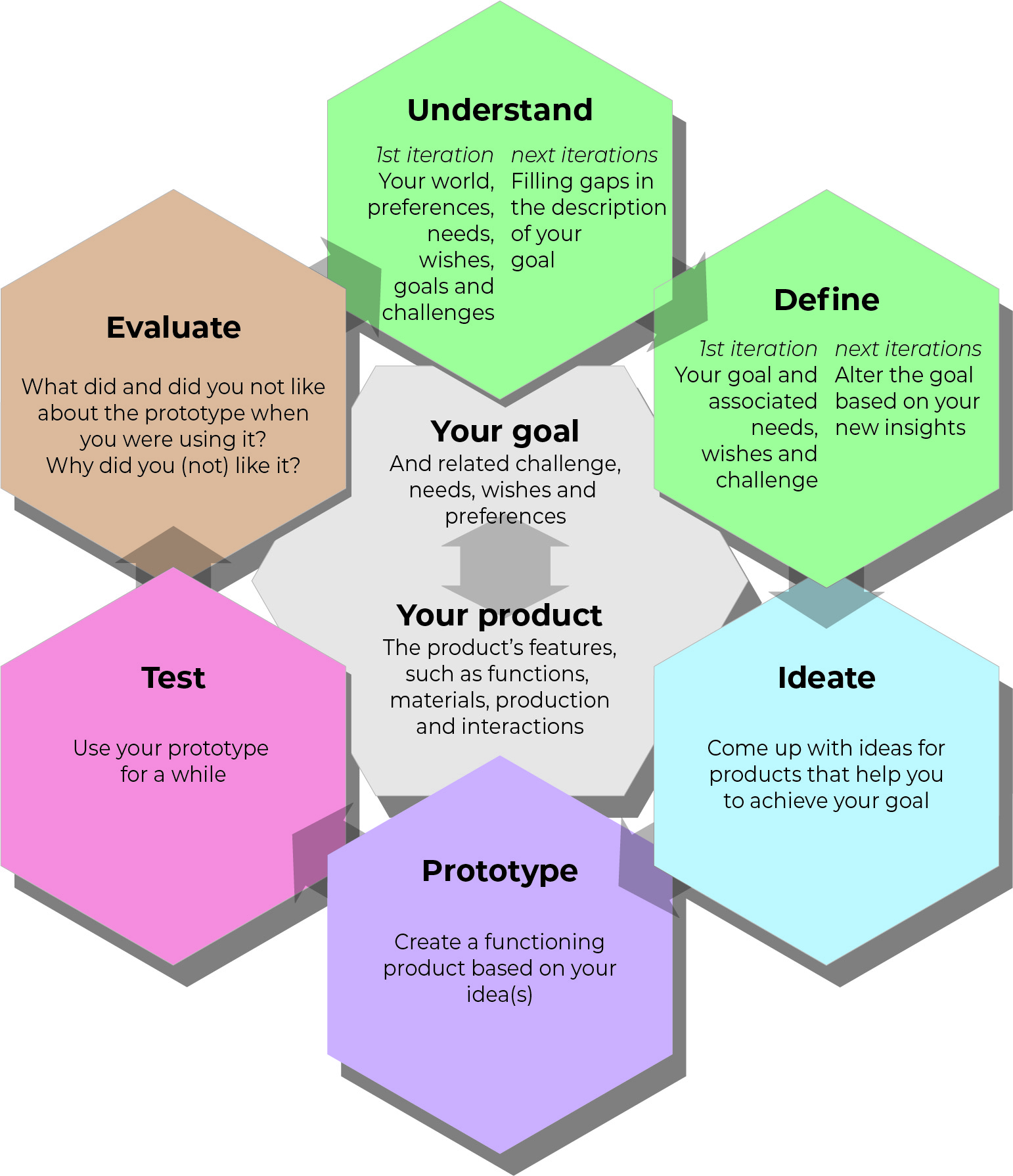}
\caption{The Design Your Life process that resulted from the first three case studies.} \label{fig5}
\end{figure}
The process presented here encompasses six design stages: Understand, Define, Ideate, Prototype, Test and Evaluate. They revolve around the development of goals and products.

\subsection{Design Stages}
\subsubsection{Goals and Products} At the centre are the goals and products, that will develop during the process. The bidirectional arrow indicates that they influence each other, meaning that a product can also lead to the insights and development of one’s goals and vice versa.
\subsubsection{Understand} The focus here is on establishing a self-understanding (“Who am I?”, “What is the world that I live in?”), starting from the fact that a YAA and a carer already know each other, at least to a certain degree. So, the goal is to make the insights into one-self and one’s world more explicit. Not for a designer or researcher, but for oneself, to unlock design ideas and make them actionable.
\subsubsection{Define} This phase focuses on defining a specific purpose for the solution (“What should the technology support?”) as relative to a larger life goal that a YAA and caregiver define as well. 
\subsubsection{Ideate} The ideation phase is about broadening the solution-space. It is aimed at stimulating creativity to come up with possible (non-obvious) directions for a solution.
\subsubsection{Prototype} In this phase products are acquired or realised. Depending on the available resources, possibilities and affinity with technologies, choices are made how to realise product(s). These conditions influence the extent to which technology can be adapted for its own application. Most people will be able to use technology, but fewer will (be able to) configure, modify or even create technologies themselves. This observation is visualised in the form of what is called here the ‘Pyramid of technological personalisation’ (Figure 6). This was inspired by the four levels of creativity described by Sanders and Stappers \cite{sanders_co-creation_2008}.
\subsubsection{Test} Here, products are being used in everyday life. It will be determined if and how it works and to what extent it contributes to the independence of the YAA.
\subsubsection{Evaluate} It was chosen to create an extra phase to evaluate the whole process. It is aimed at gaining insights to refine self-understanding.
\begin{figure}
\includegraphics[width=\textwidth]{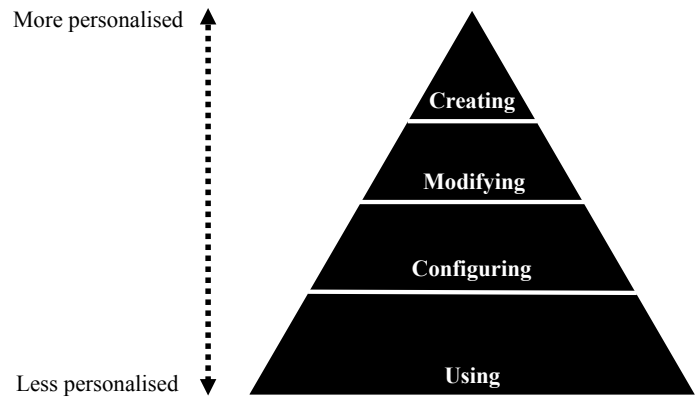}
\caption{Pyramid of technological personalisation: most people use technology, fewer configure, modify or create technology.} \label{fig6}
\end{figure}

\section{Conclusion}
This paper offers the first version of the Design Your Life-method: a novel approach that is aimed at increasing the effectiveness of technologies that support young autistic adults in independent living. It is based on three case studies and inspired by phenomenology synthesised into four core principles: ‘focus on experience’, ‘user-initiated design’, ‘action-oriented tinkering’ and ‘off-the-shelf technologies’. These principles will put the design process firmly into both the lived experience of young autistic adults as well as being integrated into the daily practice of care. \par
The added value of this method lies not so much in the ‘designerly approach’, but mainly in its phenomenological foundations. This will be given a more prominent role in the next iterations of the method. Furthermore, the extent to which the findings are generic or unique will be explored in more detail. The question is to what extent the method itself should be adaptable to meet users' preferences (i.e. Should the DYL-method itself be personalised?). \par
As described in the last part of this paper, users of the method will only be able to personalise technology to their own needs to a certain extent. So, the effectiveness of the method will probably depend on the willingness and capacities of the users. Another important factor is that applying the proposed method will have an impact on the caregivers and organisations who provide care to YAA. To anticipate a better collective understanding and support throughout the organisation, multi-stakeholder co-reflection sessions will be organised \cite{tomico_designers_2011}, in which the results of the case studies are interpreted from the different perspectives. This gives a more robust reflection of the insights: it identifies the challenges faced and forms the starting point for the co-design case studies that will follow. \par
The case studies were conducted between March and November 2020. It was therefore necessary to anticipate the Covid-19 circumstances. For example, shorter and far fewer face-to-face co-design gatherings could be organised. This may have influenced the outcome of the case studies. On the other hand, it also sparked inspiration to explore other ‘modes of cooperation’. For example, Johansen’s Time-Space Matrix \cite{johansen_groupware_1988} offered inspiration for this: remote and/or asynchronous co-design will be further explored.

\subsubsection*{Acknowledgements.}We thank all participants and organisations who took part in this research. Thanks to Nathalie Overdevest, Industrial Design Engineering student from the University of Twente, for her contributions to the development and visual design of the DYL-model. We also thank Laura van den Berg, Brian Schipper and Jasmijn Sagel, Industrial Design Engineering students from the University of Twente, who designed the models and realised the prototypes for the case studies. This research is funded by the Dutch Taskforce for Applied Research SIA (RAAK.PRO.03.045) and the Dutch Research Council (Aut.19.007).

%
%
\printbibliography
\end{document}